\documentclass[sigconf]{acmart}  
\AtBeginDocument{%
  }

\setcopyright{acmlicensed}
    
\copyrightyear{2026}
\acmYear{2026}
\setcopyright{cc}
\setcctype{by}
\acmConference[MSR '26]{23rd International Conference on Mining Software Repositories}{April 13--14, 2026}{Rio de Janeiro, Brazil}
\acmBooktitle{23rd International Conference on Mining Software Repositories (MSR '26), April 13--14, 2026, Rio de Janeiro, Brazil}
\acmPrice{}
\acmDOI{10.1145/3793302.3793309}
\acmISBN{979-8-4007-2474-9/2026/04}
\usepackage{stfloats}
\usepackage{multirow}

\begin{document}

\title{InEx-Bug: A Human Annotated Dataset of Intrinsic and Extrinsic Bugs in the NPM Ecosystem}

\author{Tanner Wright}
\email{tanner.wright@ubc.ca}
\affiliation{%
  \institution{University of British Columbia}
  \city{Kelowna}
  \state{British Columbia}
  \country{Canada}
  }
\author{Adams Chen}
\email{achychen@student.ubc.ca} 
\affiliation{%
  \institution{University of British Columbia}
  \city{Kelowna}
  \state{British Columbia}
  \country{Canada}
}

\author{Gema Rodr\'iguez-P\'erez}
\email{gema.rodriguezperez@ubc.ca} 
\affiliation{%
  \institution{University of British Columbia}
  \city{Kelowna}
  \state{British Columbia}
  \country{Canada}
}

\renewcommand{\shortauthors}{Wright et al.}

\begin{abstract}
Understanding the causes of software defects is essential for reliable 
software maintenance and ecosystem stability. However, existing bug datasets do not distinguish between issues originating within a project from those caused by external 
dependencies or environmental factors. In this paper we present InEx-Bug, a manually annotated dataset of 377 GitHub issues from 103 NPM repositories, categorizing issues as Intrinsic (internal defect), Extrinsic (dependency/environment issue), Not-a-Bug, or Unknown. 
Beyond labels, the dataset includes rich temporal and behavioral metadata such as maintainer participation, code changes, and reopening patterns. Analyses show Intrinsic bugs resolve faster (median 8.9 vs 10.2 days), are close more often (92\% vs 78\%), and require code changes more frequently (57\% vs 28\%) compared to Extrinsic bugs. While Extrinsic bugs exhibit higher reopen rates (12\% vs 4\%) and delayed recurrence (median 157 vs 87 days). The dataset provides a foundation for further studying Intrinsic and Extrinsic defects in the NPM ecosystem. 
\end{abstract}



\maketitle

\section{Introduction}
Modern software systems increasingly rely on large and dynamic ecosystems of third-party packages and libraries. While this dependency structure accelerates development and promotes code reuse, it also introduce risks as a single defect can propagate through transitive dependencies, breaking functionality across thousands of downstream projects~\cite{cogo2019empirical,dusing2022analyzing, jayasuriya2025extended,kikas2017structure}. In the NPM ecosystem, which host over 3 million packages~\cite{decan2018impact}, 
maintainers must regularly determine whether reported issues originate from their own codebase (Intrinsic bugs~\cite{rodriguez2020how}) or arise due to external factors such as dependency updates, environment changes, or upstream API modifications (Extrinsic bugs~\cite{rodriguez2020how}). Distinguishing between these types is often non-trivial yet critical for effective maintenance as Intrinsic and Extrinsic bugs differ significantly in the size of the code changes required for their resolution~\cite{rodriguez2022watch}.

Moreover, prior research shows that approximately 33.8\%~\cite{herzig2013feature} to 40\%~\cite{rodriguez2016bugtracking} of issues in bug tracking system are mislabeled, meaning they are not true bugs but user errors, questions, or feature requests. Such mislabeling threatens the reliability of datasets and can negatively affect the development and evaluation of automated tools~\cite{herzig2013feature}. To date, the only dataset containing Intrinsic and Extrinsic labels was curated in~\cite{rodriguez2022watch},  however, it was limited to commit-level data in a single project (OpenStack), leaving open questions about characteristics of bug origins  at the issue-tracking level across diverse ecosystems like NPM. 

To address these gaps, we introduce \textbf{InEx-Bug~\cite{wright2025inexbug}}, a manually annotated dataset of 377 GitHub issue from 103 NPM repositories. Each issue is classified as Intrinsic, Extrinsic, Not-a-Bug, or Unknown based on analysis of issue descriptions, discussions, linked code changes, and timeline events. Unlike existing bug datasets that focus on either code-level or text-level labels~\cite{ferenc2018public,just2014defects4j,saeidi2025bug,herbold2022fine,karampatsis2020often}, InEx-Bug combines the bug origin labels with rich contextual and behavioral metadata, including temporal metrics,  participation data, and code changes characteristics.

By offering a fine-grained, human-validated perspective on bug origins and maintainer behaviors in NPM, InEx-Bug can support research on automated bug classification, dependency risk analysis or maintainer decision-making. In doing so, it contributes foundational data for advancing our understanding of how Intrinsic and Extrinsic defects emerge, propagate, and are managed in large-scale software ecosystems.

\section{Dataset Construction Methodology}


\subsection{Data Filtering and Sampling } The dataset used in this study originates from Saeidi et. al \cite{saeidi2025bug}, which comprises a larger collection of approximately 30,000 GitHub issues and pull requests (PRs) from 500 of the most depended upon NPM packages. To focus on issue reports relevant to Intrinsic and Extrinsic bug analysis, all PRs were removed from the starting 30,000 bug reports, resulting in 23,682 issues.

These remaining issues served as input for our data collection process. From the 23,682 bug reports, we determined that a sample of 377 issues provides 95\% confidence level with $\pm$5\% margin of error for estimating population proportions. Accordingly, we randomly selected 377 issue URLs from the filtered dataset for data collection.

\subsection{Metadata Extraction} 
Metadata collection was performed in October 2025 using a custom Python script that retrieves comprehensive metadata utilizing the GitHub REST API v3. For each issue, we collected core textual metadata (title, body, state, timestamps, labels, assignees, milestone), associated comments and authors, complete timeline events (e.g., closed, reopened, referenced, cross-referenced), and when applicable, detailed PR or commit information, including code changes, reviewers, and merge or commit timestamps. A full list of all collected attributes is documented in our replication package~\cite{wright2025inexbug}.

To identify issues that were closed through code changes (i.e., via PR or direct commits)~\cite{kalliamvakou2014promises}, we implemented a multi-strategy detection approach. We examined the issue's closing event for direct PR references, then searched timeline events for cross-references linking to PRs or commits. For each candidate, we verified that the code change occurred after issue creation and that the merge/commit timestamp fell within 7 days of issue closure. While GitHub automatically 
closes issues within seconds when PRs use closing keywords, this 7-day window also captures workflows where maintainers manually close issues after verifying fixes in production or following  deployment cycles. 
We note that this threshold is a design choice that balances capturing legitimate code-based closures with minimizing unrelated commits and might threat the results of this study. But can be adjusted by other researchers depending on project activity. Issues closed by direct commits are tracked separately in the \texttt{closing\_commit} field.

The resulting dataset uses JSONL format for efficient processing. Each issue record contains: \textbf{(1)} core metadata (title, body, state, timestamps, labels, assignees, milestone); \textbf{(2)} discussion data (all comments with author associations, comment counts);
\textbf{(3)} derived temporal metrics (time to first response, time to close, and other timeline-based measures); \textbf{(4)} participation metrics (unique commenters, maintainer involvement, and roles); \textbf{(5)} closure patterns (reopen events, time to reopen); \textbf{(6)} code-based closure details when applicable (PR or commits identified from cross-reference timeline events, including code changes, reviewers, and merge timestamps); and \textbf{(7)} manual classification labels (Intrinsic, Extrinsic, Not-a-Bug, and Unknown). See~\cite{wright2025inexbug} for the full list definitions.


Maintainers are defined as users whose author association with the project is OWNER, MEMBER, COLLABORATOR, or CONTRIBUTOR. Two automated bot accounts (\texttt{stale[bot]} and \texttt{vue-bot}) were identified and flagged through manual analysis in our dataset for exclusion from human activity analyses (12 issues, 3.2\% of the dataset). Closure rates reflect human maintainer activity by measuring the percentage of issues closed through human actions rather than automated processes. In section \ref{sec:Results}, all 12 bot-closed issues are fully excluded from all timing analyses.


\begin{figure*}[!t]
    \centering
    \includegraphics[width=0.8\textwidth]{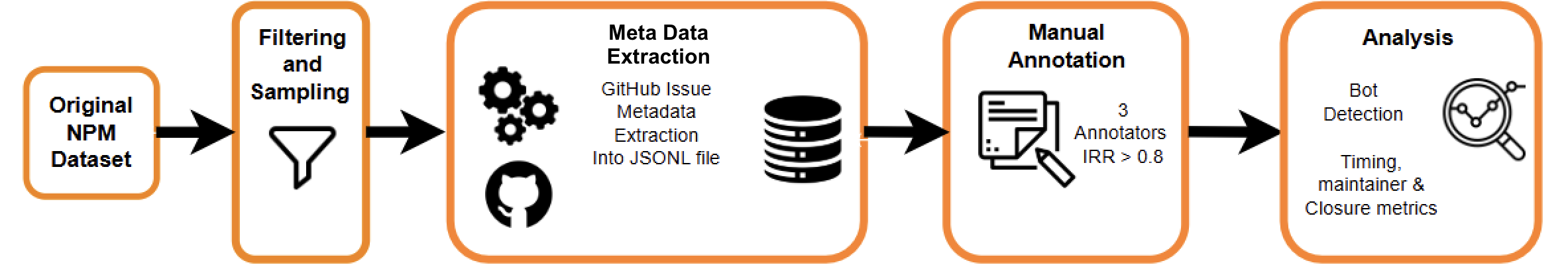}
    \caption{ Pipeline consisting of (1)~ original data, 
    (2)~filtering and sampling, (3)~metadata extraction (4)~manual annotation, and (5)~data analysis}
    \Description{A image showing the Pipeline for constructing the InEx-bug dataset.}

    \label{fig:pipeline}
\end{figure*}







\subsection{Annotation Process} 



\textbf{Bug Classification} Each issue was manually examined and classified into one of four categories: \texttt{Intrinsic}, \texttt{Extrinsic}, \texttt{Not-a-Bug}, or \texttt{Unknown}, following a structured rubric refined through multiple annotation iterations. This process ensured consistent interpretation of root causes across repositories and accurately distinguished internal project defects from issues arising from external dependencies or the environment. The classification reflects the source and nature of the underlying problem, from the perspective of the repository receiving the bug report. For instance, an issue could be classified as \textit{Extrinsic} even if resolving it required a local code change, provided the defect originated externally. Table~\ref{tab:classification-framework} summarizes the operational criteria used during annotation.

\begin{table}[h]
\centering
\caption{Definitions for each classification category.}
\label{tab:classification-framework}
\begin{tabular}{p{0.15\linewidth}p{0.8\linewidth}}
\toprule
\textbf{Label} & \textbf{Definition and common types} \\
\midrule

\textbf{Intrinsic} & The bug originates within the package to which the report belongs~\cite{rodriguez2020how}. Typical cases include regressions bugs, 
database or logic errors, or any defect introduced in the project's own code.\\ 

\textbf{Extrinsic} & The bug is caused by changes or incompatibilities external to the project, such as dependency upgrades, operating system differences, or browser-specific behavior. While the project’s code functioned correctly under previous conditions, it fails following these external modifications~\cite{rodriguez2020how}. This includes issues such as dependency incompatibilities (“bug in dependency X”), OS-specific errors, or adaptation required after another project’s update. \\

\textbf{Not-a-Bug} & The report does not represent an actual defect in the software. These include 
cases such as test or documentation issues, enhancement requests, questions, license clarifications, configuration misuse, or bugs reported in the wrong repository.\\

\textbf{Unknown} & Reports lacking sufficient context to determine the root cause.  \\
\bottomrule
\end{tabular}
\end{table}

\textbf{Manual annotation process:} The dataset was manually annotated using a rigorous, multi-stage process. Two annotators (the first and third authors) jointly labeled 161 issues over five iterative rounds, discussing disagreements after each round to ensure consistent interpretation of the classification rubric and accurate identification of bug report types. Inter-rater reliability (IRR), measured using Cohen’s kappa \cite{cohenkappa}, improved with each round, reaching $\kappa > 0.80$ in the final round. Once substantial agreement was established, the remaining 216 issues were independently labeled by the first author using the finalized definitions. To validate both the independent annotations and the classification rubric, a separate researcher classified 234 of the 377 issues. Any discrepancies were resolved through discussion, and cases with persistent ambiguity were flagged and collectively reviewed by the research team to reach consensus.



\textbf{Limitations:} The dataset captures the state of repositories as of October 2025, and subsequent changes may not be reflected. Metrics like “time to close” or “maintainer response rate” may be influenced by workflow factors rather than defect complexity. While the dataset includes a range of library sizes and activity levels, observed patterns (e.g., Extrinsic bug proportion, maintainer engagement) may not generalize to other ecosystems or industrial contexts. Statistical comparisons are descriptive, and the dataset is intended to provide a high-quality annotated resource rather than support hypothesis testing.


\section{Data Analysis}
\label{sec:Results}

\textbf{\underline{Distribution}}:
Table~\ref{tab:class-summary} shows the classification distribution. Not-a-Bug 
reports comprised 58.9\% (222 issues), followed by Intrinsic bugs at 27.9\% 
(105 issues), Extrinsic bugs at 9.8\% (37 issues), and Unknown at 3.4\% (13 issues). 
Among actual bug reports, Intrinsic bugs outnumbered Extrinsic bugs by approximately 3 times.

\textbf{\underline{Response Times}}:
Table~\ref{tab:class-summary} presents key temporal metrics by classification. The median time to first response was 8.6 hours for Intrinsic bugs, 12.4 hours for Extrinsic bugs, 6.2 hours for Not-a-Bug reports, and 28.4 hours for Unknown issues. Maintainer participation rates were high across all categories: 86.7\% 
for Intrinsic, 89.2\% for Extrinsic, and 80.6\% for Not-a-Bug reports. Comment activity was similar across bug types, with Intrinsic issues averaging 5.4 comments, Extrinsic issues averaging 5.5 comments, and Not-a-Bug reports averaging 4.0 comments per issue.

\begin{table*}[t]
\centering
\caption{Issue characteristics by class. All timing metrics exclude bot-closed issues; medians are reported for time values.}
\label{tab:class-summary}
\renewcommand{\arraystretch}{1.3}
\resizebox{\textwidth}{!}{%
\begin{tabular}{lrrrrrrrrrrrrr}
\toprule
\textbf{Class} &
\textbf{Count} &
\textbf{Closed \%} &
\textbf{Med. Close (d)} &
\textbf{Med. 1st Resp (h)} &
\textbf{Mean Comments} &
\textbf{Mean Participants} &
\textbf{Maint.\ Resp.\%} &
\textbf{Reopen \%} &
\textbf{Code Closure \%} &
\textbf{Mean Lines} &
\textbf{Med. Lines} &
\textbf{Mean Files} &
\textbf{By PR/Commit \%} \\
\midrule
\textbf{Intrinsic} & 105 (27.9\%) & 92.4 & 8.9 & 8.6 & 5.4 & 3.4 & 86.7 & 3.8 & 56.7 & 369 & 51 & 4.6 & 43.3 / 13.4 \\
\textbf{Extrinsic} & 37 (9.8\%) & 78.4 & 10.2 & 12.4 & 5.5 & 3.6 & 89.2 & 11.8 & 28.1 & 34 & 11 & 3.2 & 18.8 / 9.4 \\
\textbf{Not-a-Bug} & 222 (58.9\%) & 93.2 & 0.7 & 6.2 & 4.0 & 2.9 & 80.6 & 1.9 & 7.5 & 146 & 49 & 4.1 & 5.6 / 1.9 \\
\textbf{Unknown} & 13 (3.4\%) & 76.9 & 2.1 & 28.4 & 2.1 & 2.2 & 46.2 & 0.0 & 0.0 & --- & --- & --- & --- / --- \\
\bottomrule
\end{tabular}%
}
\vspace{0.5ex}
\end{table*}

\begin{figure}[!t]
  \centering
  \includegraphics[width=0.45\textwidth]{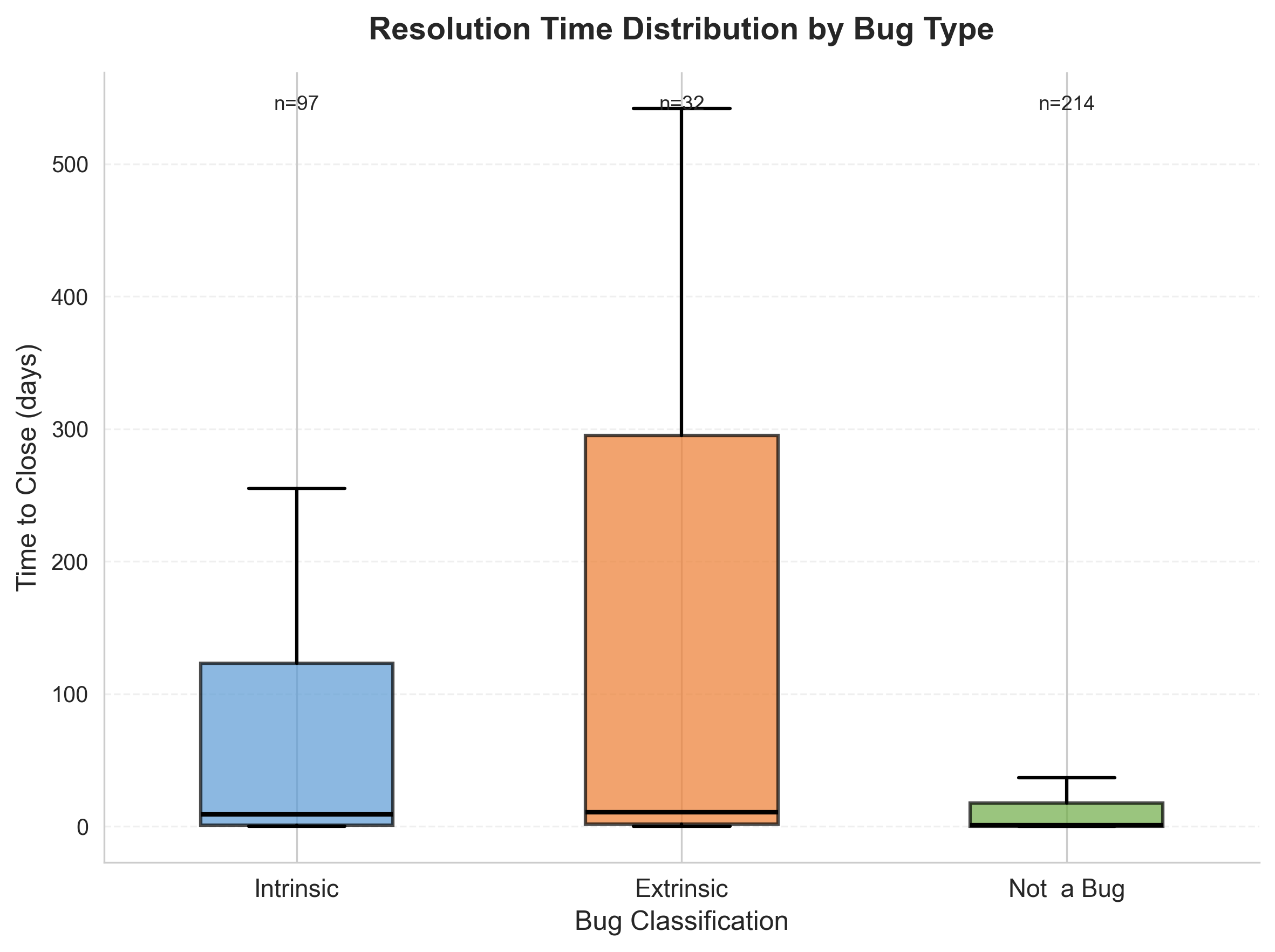}
  \caption{Distribution of time to close by bug classification.}
  \Description{A box plot showing the distribution of issue closure times by category.}
  \label{fig:timetoclose}
\end{figure}

\textbf{\underline{Closure Patterns}}:
Intrinsic bugs closed faster than Extrinsic bugs (median 8.9 vs 10.2 days), and had higher closure rates (92.4\% vs. 78.4\%). Not-a-Bug reports closed rapidly (median 0.7 days, 93.2\% rate). Figure~\ref{fig:timetoclose} 
illustrates the time-to-close distributions, showing that while medians differ 
moderately, both Intrinsic and Extrinsic bugs exhibit long-tailed distributions relative to Not-a-Bug bug reports where Intrinsic did not go past 275 days and Extrinsic went past 500. 

\textbf{\underline{Reopening}}:
Reopen events occurred in 3.8\% of Intrinsic bugs, 11.8\% of Extrinsic bugs, and 1.9\% of Not-a-Bug reports. For reopened issues, the median time from initial closure to first reopen was 87 days for Intrinsic bugs and 157 days for Extrinsic bugs.

\textbf{\underline{Code Changes}}:
Code-based closed issues resolved via merged pull requests or direct commits within seven days of closure, occurred in 21.2\% of all issues. Table~\ref{tab:class-summary} 
shows the breakdown by classification. Intrinsic bugs were most frequently 
closed via code (56.7\% of closed issues), compared to 28.1\% for Extrinsic 
bugs and 7.5\% for Not-a-Bug reports. The size of code changes also differ. Fixes for Intrinsic bugs averaged 369 lines changed (median 51) across 4.6 files, whereas Extrinsic adaptations involved far fewer modifications, averaging 34 lines (median 11) across 3.2 files.

\begin{figure}[!t]
  \centering
  \includegraphics[width=0.45\textwidth]{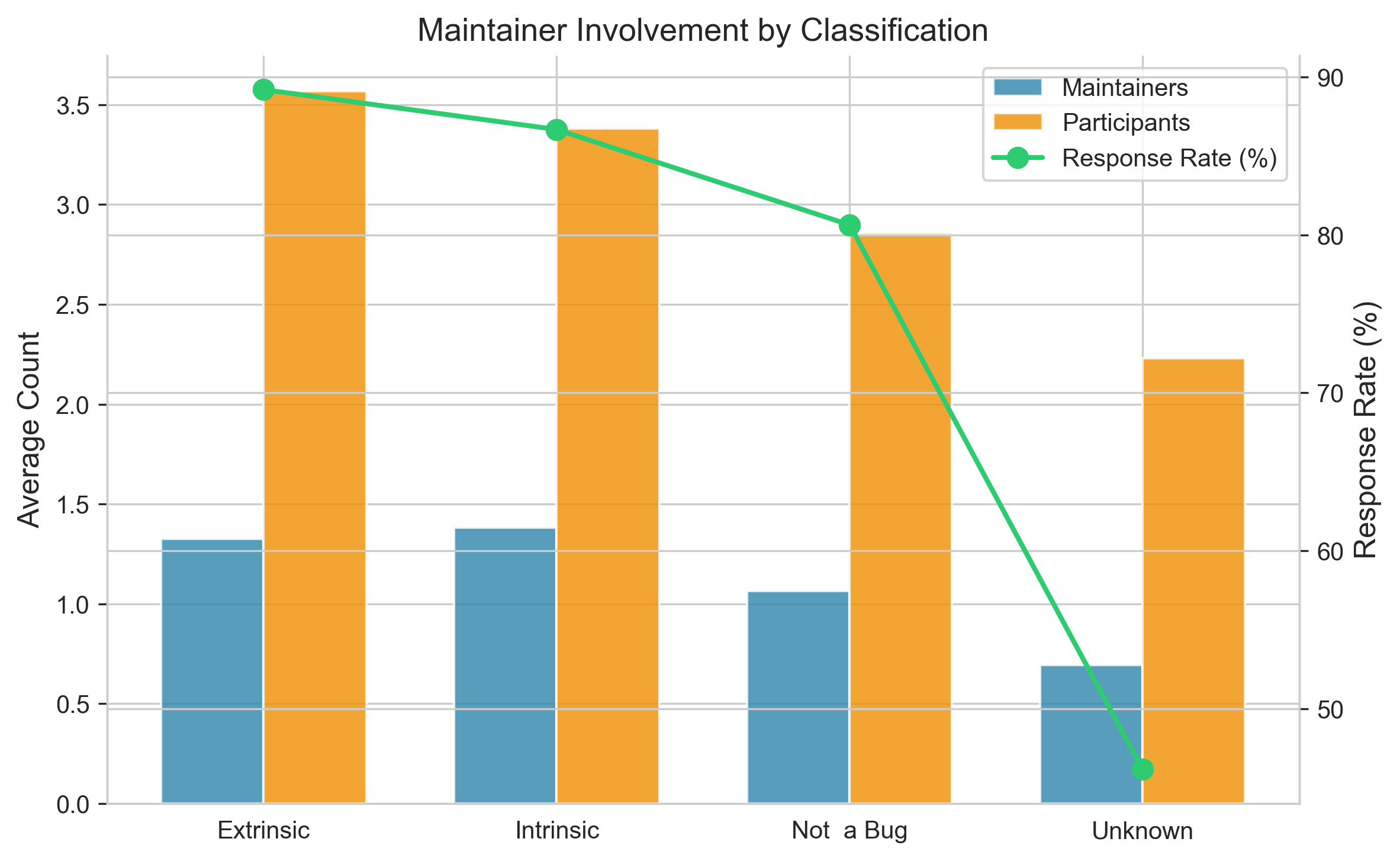}
  \caption{Maintainer involvement across classes, showing response rate and maintainer participation ratio.}
   \Description{A bar chart comparing maintainer response rates and participation ratios across issue classes.}
  \label{fig:maintainers}
\end{figure}

\textbf{\underline{Maintainer Involvement}}:
Figure~\ref{fig:maintainers} illustrates maintainer engagement patterns across 
classifications. Response rates were consistently high for actual bugs, with 86.7\% of Intrinsic and 89.2\% of Extrinsic issues receiving at least one maintainer response, compared to 80.6\% for Not-a-Bug reports. The average number of maintainers per issue was similar across bug types (1.3-1.4 
maintainers), as was the overall discussion participation (3.4-3.6 total 
participants for bugs, 2.9 for Non-bugs). These patterns indicate that maintainers engage with both Intrinsic and Extrinsic bugs at comparable levels during the discussion phase. 

\section{Discussion and Implications}

The analysis of 377 human annotated issues reveals distinct behavioral signatures between Intrinsic, Extrinsic, and Non-bug reports.  
While Intrinsic and Extrinsic bugs share similar levels of maintainer engagement and discussion depth, their underlying life cycles differ in meaningful ways that reflect the broader structure of open-source maintenance ecosystems.

\textbf{\underline{Intrinsic vs. Extrinsic Defect Dynamics}}:
Intrinsic bugs, tend to elicit direct code modifications and faster resolution cycles.  
Their shorter median closure time (8.9 days) and higher code-based closure rate (56.7\%) demonstrate that maintainers can quickly identify and patch issues under their control.  
By contrast, Extrinsic bugs, often require coordination across project boundaries and reappear long after initial resolution.  
The higher reopen frequency (11.8\%) and longer delay before reopening (157 days) suggest that dependency-related regressions might propagate slowly and unpredictably, consistent with the decentralized nature of the NPM ecosystem.

\textbf{\underline{Non-Bug Reports and Maintainer Workload}}:
58.9\% of all analyzed issues were classified as Not-a-Bug, underscoring a major portion of maintainer effort devoted to user support, misconfiguration clarification, and documentation feedback. Although such issues are resolved quickly (median 0.7 days), they consume attention that could otherwise be allocated to technical defects. This finding highlights a constant need of attention in open-source projects, specifically with NPM as it can be very confusing with packages constantly depending on other packages.

\subsection{Dataset Utility and Research Applications}
The InEx-Bug dataset, provides a valuable resource for both maintainers and tool designers, offering reliable insights that can guide more effective bug management, more effective tooling, and empirical studies of software ecosystems.


\textbf{Improved community reporting guidelines.} The dominance of Non-bug reports points to the need for more defined issue templates, validation steps, or automated diagnostics. Future work can investigate strategies to reduce the number of Not-bug reports in the NPM ecosystem. Our dataset can be used as a starting point to help identify patterns to guide users toward providing higher-quality, actionable reports.

\textbf{Understanding Maintainer Behaviors}
Utilizing the InEx-Bug metadata collected can allow for studies of maintainer behavior patterns and response dynamics across repositories. 
Researchers can investigate how a bugs origin influences different types of maintainer attention and analyze participation patterns across different types of issues. 

\textbf{Analyzing Ecosystem Dependencies}
InEx-Bug supports empirical studies of dependency-induced issues and their propagation through package networks. Researchers can examine patterns in Extrinsic bug manifestation, investigate the relationship between dependency updates and issue reopening, and quantify the coordination effort required when fixes span multiple projects or ecosystems.

\section{Related Work}

\textbf{Bug origins.} The foundational work by Rodr\'iguez-P\'erez et al.~\cite{rodriguez2020how} established the theoretical framework for distinguishing Intrinsic from Extrinsic bugs in commit-level data. They defined Extrinsic bugs as defects arising from factors not recorded in version control systems, such as dependency updates or requirement changes. Follow-up work~\cite{rodriguez2022watch} showed that 9–21\% of bugs in OpenStack were Extrinsic and that conflating bug types degraded the accuracy of Just-In-Time defect prediction models. However, both studies were limited to commit-level analyses within individual projects, leaving open questions about how bug origins manifest at the issue-tracking level across ecosystems. Tools such as BuggIn~\cite{ferenc2018public} proposed NLP-based automated classification for Intrinsic bugs but lacked manually verified labels and did not explicitly address Extrinsic bugs. Our work extends this line of research by providing the first manually annotated dataset distinguishing Intrinsic from Extrinsic bugs at the issue level across diverse NPM packages.

\textbf{Bug report datasets.} Herzig et al.~\cite{herzig2013feature} found that approximately 33\% of reports marked as bugs were actually misclassifications (features, refactorings, or non-issues), highlighting the need for manual verification. While datasets such as Eclipse and Mozilla~\cite{lamkanfi2013eclipse} provide rich issue-tracking histories, they do not classify bugs by origin. More recent datasets like AndroR2~\cite{andror2021} focus on reproducing mobile app bugs, whereas BugsPHP and BugsJS target language-specific issues for automated program repair. In contrast, InEx-Bug explicitly distinguishes internal defects from dependency-induced issues and provides comprehensive behavioral metadata essential for studying ecosystem maintenance dynamics.



\section{Conclusion}
This paper presents InEx-Bug, a manually annotated dataset of 377 GitHub issues classified as Intrinsic, Extrinsic, or Non-bug through a rigorous multi-round annotation process with consensus and reliability checks. The dataset integrates both textual and behavioral metadata, enabling detailed analysis of project maintenance dynamics. Our findings reveal notable differences between Intrinsic and Extrinsic bugs, emphasizing the importance of distinguishing bug origins in empirical studies. This paper provides reusable scripts for reproducing quantitative results and extending the dataset to new repositories or ecosystems~\cite{wright2025inexbug}. 

\section{Acknowledgements}

We would like to thank the Natural Sciences and Engineering Research Council of Canada (NSERC) Alliance (RGPIN-2022-03265) for funding this research.
\bibliographystyle{ACM-Reference-Format}
\bibliography{software.bib}

\end{document}